\documentclass[twocolumn,english,aps,superscriptaddress,prb]{revtex4-2}

\usepackage[colorlinks=true,allcolors=blue]{hyperref}
\usepackage{array}
\usepackage[latin9]{inputenc}
\usepackage{amsmath}
\usepackage{amssymb}
\usepackage{graphicx}
\usepackage{babel}
\usepackage{mathrsfs}
\usepackage{amsfonts}
\usepackage{epstopdf}
\usepackage{multirow}
\usepackage{color}
\usepackage{natbib}
\usepackage{bm}
\usepackage{upgreek}
\usepackage{xcolor}
\usepackage{verbatim}

\begin{document}

\title{Dynamic critical exponents as an emergent property at interacting topological quantum critical points}
\author{Fan Yang}
\affiliation{School of Physics, Southeast University, No.2 SEU Road, Nanjing, China, 211189}
\date{\today}

\author{Zheng-Cheng Gu}
\affiliation{Department of Physics, Chinese University of Hong Kong, Sha Tin, Hong Kong, China}

\author{Fei Zhou}
\affiliation{Department of Physics and Astronomy, University of British Columbia, 6224 Agriculture Road, V6T 1Z1}

\begin{abstract}
In standard studies of quantum critical points (QCPs), the dynamic critical exponent $z$ is introduced as a fundamental parameter along with global symmetries to identify universality classes. 
Often, the dynamic critical exponent $z$ is set to be one as the most natural choice for quantum field theory representations, which further implies emergence of higher space-time symmetries near QCPs in many condensed matter systems. 
In this article, we study a family of topological quantum critical points (tQCPs) where the $z=1$ quantum field theory is prohibited in a fundamental representation by a protecting symmetry, resulting in tQCPs with $z=2$. We further illustrate that when strong interactions are properly taken into account, the stable weakly interacting gapless tQCPs with
$z=2$ can further make a transition to another family of gapless tQCPs
with dynamic critical exponent $z=1$, without breaking the protecting symmetry. Our studies suggest that dynamic critical exponents, as well as the degrees of freedom in fermion fields, can crucially depend on interactions in topological quantum phase transitions;
in tQCPs, to a large extent, they are better thought of as {\em emergent properties}. 
\end{abstract}

\maketitle

\section{Introduction}
In the past two decades, the concept of symmetry-protected topological (SPT) states has been intensively studied \cite{SPT1,SPT2,SPT3,Schnyder08,Kitaev09}.
One of the most fascinating topics is how one SPT phase transitions to another topologically distinct one \cite{Dunghai2017, Dunghai2019,Ji20}. Although quantum critical phenomena in the standard order-disorder paradigm have been quite well understood  \cite{Landau,Fisher74,Justin,Sachdev}, the topic of topological quantum critical points (tQCP) has not been adequately studied and requires further exploration. Interestingly, it has been shown that even supersymmetry (SUSY) may naturally emerge in surface topological phase transitions \cite{SUSY1}.

The universality of conventional QCPs is often uniquely determined by the symmetry group $G$ of the Hamiltonian and spatial dimensionality. Interactions can play a role in quantum critical phenomena and even drive a transition,  but the universality class itself does not rely on details of interactions or their explicit forms if the interactions are local, following the well-known Wilsonian approach to critical phenomena \cite{Wilson71}.
This simplicity of the conventional approach to QCPs is largely due to the fact that the quantum field theory representations, especially the fundamental ones (i.e. with minimum degrees of freedom) of QCPs are fully set by symmetries. The most relevant operators one can construct in these representations generally lead to a dynamic critical exponent $z=1$. This often results in quite restricted families of scale-conformal invariant fixed points available for a given symmetry group $G$ of a broad class of condensed matter systems. 

Quantum critical phenomena with $z=2$ are much rarer; they are mainly found at transitions such as those to ferromagnetic states and Lifshitz or Lifshitz-like transitions with emergent non-relativistic Galilean invariance \cite{Lifshitz60,Fisher89}. In the cases known to us so far, QCPs with dynamic critical exponent $z=2$ (such as the ones stated above) can be very well isolated from the $z=1$ classes. In addition, the invariant group of critical states is simply $G$, the symmetry group of the Hamiltonian, or a subgroup of $G$ in the case of multiple stages of spontaneous symmetry breaking. On the other hand, the $z=2$ classes with charge-conjugation symmetry are often associated with Lifshitz multicritical points with higher degrees of fine-tuning \cite{Rokhsar88,Fradkin04}. 

In contrast, the complexity of tQCPs is at least there-fold. First, apart from protecting symmetries $G_p$ (the subscript $p$ refers to protecting), topology also plays an important role. This adds a new dimension to tQCPs compared to conventional QCPs. A generic tQCP represents an interplay between topology, symmetry, and interaction. The second complexity is that tQCPs often display puzzling emergent symmetries and the invariant group of a quantum gapless state at tQCPs can be larger than the protecting symmetry $G_p$ \cite{Zhou22,Zhou23}. This potentially opens up  more discussions on 't Hooft anomalies near tQCPs, and their connections to higher dimensional gapped SPT surfaces with gauge anomalies \cite{Senthil13, Wen13,Bi19,Zhou24}. 
The third complexity is that {\em quantum field theory representations of tQCPs themselves} are not uniquely set by the topology and symmetry, {\em but can further rely crucially on interactions}. This rather surprising aspect of tQCPs is the focus of this paper. In this sense, interactions can play a much more deciding and even a paramount role at tQCPs. In this paper, we only focus on local interactions, so that the effective field theories are local.

An immediate consequence of the last point is that even the dynamic critical exponent $z$ and the low-energy degrees of freedom in quantum fields can rely on interactions. Therefore, both of them are better thought of as emergent parameters.
This is very different from the conventional paradigm of QCPs where both the dynamic critical exponent $z$ and the degrees of freedom are usually introduced as independent input parameters to define universality classes although the later quantity has to be consistent with the symmetry group $G$.

\section{Topological quantum phase transitions in superconductors with $SU(2)$ symmetry}
\begin{figure}
    \centering
    \includegraphics[width=\columnwidth]{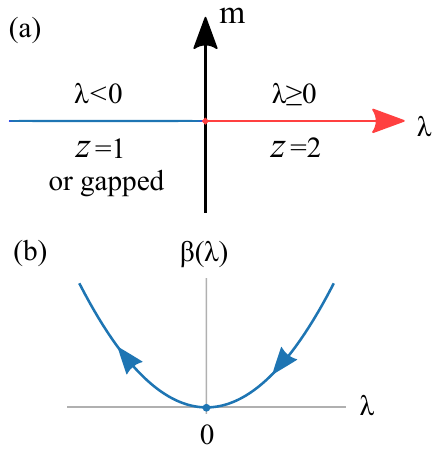}
    \caption{(a) Phase diagram of 2D $d+id$ superconductors. $m$ is the effective chemical potential. $m>0$ and $m<0$ corresponds to topological and nontopological phases. Horizontal axis marks the tQCPs with $m=0$. $\lambda$ is the contact fermion-fermion interaction. For $\lambda\geq0$, we have $z=2$ gapless tQCPs, for $\lambda<0$, depending on the interaction channel, we have either gapless tQCPs with $z=1$ or a discontinuous first order phase transition (implied by gapped states) at $m=0$. (b) The infrared renormalization flow of $\lambda$ in the $\beta(\lambda)$-$\lambda$ plane. (Here, we have neglected the normal state dispersion $\epsilon_k$, whose effect is discussed towards the end of the main text.) }
    \label{phase}
\end{figure}

In this paper, we take topological superconductors of C class \cite{Schnyder08,Kitaev09} as an example to illustrate how interactions can change dynamic critical exponents. We consider the fundamental representation of C class topological superconductors with interactions. Its protecting symmetry is the spin $SU(2)$ symmetry \cite{Kennedy16}, which demands spin singlet pairing. To realize nontrivial topology, the superconductor needs to be two-dimensional (2D) and the pairing should be in the $d_{xy}+id_{x^2-y^2}$ channel. (For simplicity, we temporarily neglect the normal state dispersion $\epsilon_k$, whose effect will be discussed in detail toward the end of this paper.) As shown in Fig. \ref{phase}, a topological phase transition can be driven by the effective chemical potential $m$, where the horizontal axis of $m=0$ marks the tQCPs. $\lambda$ is the contact fermion-fermion interaction. We find that for repulsive or no interactions $\lambda\geq0$ the tQCPs have dynamic critical exponent $z=2$, while for attractive interactions $\lambda<0$, depending on the interaction channel, we may have tQCPs with dynamic critical exponent $z=1$ or gapped states at $m=0$.
In the following, we will use effective field theory to study these tQCPs and demonstrate in detail how the dynamic critical exponent changes.

Let us define the quantum field theory representation of a tQCP by the Hamiltonian $\mathcal{H}=\mathcal{H}(N^{B}_f, \lambda)$, which contains the fermion degrees of freedom $N^{B}_f$ (the superscripts $B$ stand for bulk) and interaction parameter $\lambda$. 
A tQCP separates two topologically distinct phases with different invariants $N_1$ and $N_2$.
The degrees of freedom can be determined by both the topology and protecting symmetry $G_p$,
\begin{eqnarray}
N^B_f=N^B_f(\delta N=N_2-N_1; G_p).
\end{eqnarray}
When we further restrict the theory to the minimum change of topological invariants, $N^B_f$ would be a function of the protecting symmetry only, as $\delta N_{min}$ is a function of $G_p$ itself. Intuitively, one would expect
\begin{eqnarray}
N^B_f=N^{B}_{f}[\delta N_{min}(G_p); G_p]=N^B_f(G_p).
\label{degrees1}
\end{eqnarray}

For C class topological superconductors, which are nontrivial in 2D, we have $G_p=SU(2)$ and $\delta N_{min}=4$. In the fundamental representation, we have $N^B_f=\frac{1}{2}$, i.e., one half of Dirac fermions. It forms a fermionic representation of $spin(4)=su(2)\otimes su(2)$ algebra. 

We can identify one of the $SU(2)$ subgroups with the protecting symmetry group $G_p=SU(2)$ \cite{Zhou23}. In this fundamental representation, the free Hamiltonian $H_0$ is a four-by-four Hermitian matrix operator. Spin $SU(2)$ symmetry demands $[H_0,S_{x,y,z}]=0$, where $S_x,S_y,S_z$ are the generators of spin rotation. This puts a severe constraint on the possible gradient terms in the theory. Out of the $Spin(4)$ group, there are {\em 9 symmetric} non-unity Hermitian matrix operators which all break the protecting $SU(2)$ symmetry. The key observation is that in the fundamental representation, a gradient operator can only appear along with one of these {\em 9 symmetric} Hermitian matrix operators. So the protecting symmetry $G_p=SU(2)$, within its fundamental fermonic representation, completely excludes the terms of $z=1$ class characters.

These analyses suggest the following effective bulk Hamiltonian of a 2D C class superconductor near tQCPs, which physically describes a $d_{xy}+id_{x^2-y^2}$ superconductor
\begin{eqnarray}
&\mathcal{H}=\frac12\psi^TH_0\psi+\mathcal H_I, \nonumber \\ 
&H_0=\tau_x\sigma_y d_{xy}+\tau_z\sigma_y d_{x^2-y^2}-\tau_y(\epsilon_k-m),\nonumber \\
&d_{xy}=-2\partial_x\partial_y,\quad d_{x^2-y^2}=-\partial_x^2+\partial_y^2.
\end{eqnarray}
Here, we have used Majorana operators $\psi_{1\sigma}({\bf x})=\frac1{\sqrt2}(c_{{\bf x}\sigma}+c^\dagger_{{\bf x}\sigma})$, $\psi_{2\sigma}({\bf x})=\frac1{i\sqrt2}(c_{{\bf x}\sigma}-c^\dagger_{{\bf x}\sigma})$, and $\psi^T=(\psi_{1\uparrow},\psi_{1\downarrow},\psi_{2\uparrow},\psi_{2\downarrow})$ is a four-component real fermion. $\epsilon_k$ is the normal state dispersion and $m$ is the effective chemical potential. In most of the following discussions, we will mute $\epsilon_k$ for simplicity assuming a flat band, as it does not qualitatively affect our main results. $\mathcal H_I$ is the interaction.
$\tau_{x,y,z}$ and $\sigma_{x,y,z}$ are the Pauli matrices for $\{1,2\}$ and $\{\uparrow,\downarrow\}$ indices, respectively. 
In this representation, the spin $SU(2)$ symmetry group is generated by $S_x=-\tau_y\sigma_x, \quad S_y=\sigma_y, \quad S_z=-\tau_y\sigma_z$. In addition, $\tau_x\sigma_y,\tau_y,\tau_z\sigma_y$ generate another $SU(2)$ group that commutes with the spin $SU(2)$ group. Together, they form a representation of $Spin(4)$ group.
Along with considerations of the topology of C class states, it naturally leads to a theory with chiral $d$-wave pairing and a dynamic critical exponent $z=2$.
The free Hamiltonian suggests a topological phase transition at $m=0$ when the energy gap closes at $k=0$ with quadratic band-touching. In the following, we focus on the tQCPs and set $m=0$.

We note that Lifshitz-type transitions with $z=2$ have also been pointed out to occur near tQCPs involving gapless nodal phases \cite{Yang21}. In addition, tQCPs in noncentrosymmetric systems where the dispersion is quadratic in one direction and linear in other directions have also been studied \cite{BJYang13,BJYang14}.

\section{Strong-coupling limit and the change of dynamic critical exponent}
Next, we show that when the dynamic effect of interactions is taken into account, the dynamic critical exponent at tQCPs can become $z=1$ in the strong-coupling limit.
Let us consider contact fermion-fermion interactions of the form
\begin{equation}
\mathcal H_I=\lambda \psi^T\Gamma\psi \psi^T\Gamma\psi,
\end{equation}
where $[\Gamma,S_{x,y,z}]=0$, i.e., $\Gamma\in\{\tau_x\sigma_y,\tau_y,\tau_z\sigma_y\}$.
To understand the relevance of interactions, we first perform a one-loop renormalization group (RG) calculation, which gives 
\begin{equation}
\beta(\lambda)=\Lambda\frac{d\tilde \lambda}{d\Lambda}=(d-2)\tilde\lambda+K_d\tilde\lambda^2,
\end{equation}
where $\Lambda$ is the running momentum scale, $\tilde\lambda=\lambda\Lambda^{d-2}$ and $K_d>0$. This suggests an upper critical dimension of $d=2$. C class is nontrivial in $d=2$, where we have $K_d=1/\pi$ for $\Gamma=\tau_y$ and $K_d=1/(2\pi)$ for $\Gamma=\tau_x\sigma_y,\tau_z\sigma_y$. 
The infrared RG flow is shown in Fig. \ref{phase}. 
For $\lambda>0$, i.e., repulsive interactions, $\lambda$ is irrelevant and flows to $\lambda_c=0$. We have a topological phase transition described by free field theory with dynamic critical exponent $z=2$. For $\lambda<0$, i.e., attractive interactions,  $\lambda$ is relevant and flows to $\lambda=-\infty$. This suggests a strongly interacting theory.

To better understand the strongly interacting theory with $\lambda<0$, let us introduce an emergent real scalar field $\phi$. In the imaginary time representation, the Lagrangian can be written as follows
\begin{eqnarray}
&\mathcal{L}=\mathcal{L}_\psi+\mathcal{L}_\phi+\mathcal{L}_I,\nonumber\\
&\mathcal L_\psi=\frac12\psi^T(\partial_\tau+H_0)\psi, \nonumber \\
&\mathcal L_\phi=\frac12[(\partial_\tau\phi)^2+(\nabla^2\phi)^2+M^2\phi^2], \nonumber \\
&\mathcal L_{I}=g\psi^T\Gamma\psi\phi, \label{Yukawa}
\end{eqnarray}
where $\phi$ is spinless. With a mass gap $M^2$, $\phi$ field mediates fermion-fermion interactions with $\lambda<0$ in the infrared limit. $\phi$ also represents the emergent bosonic degrees of freedom in the limit of strong coupling. The Lagrangian is invariant under spin $SU(2)$ rotation. In this case, both the fermonic and the scalar fields have dynamic critical exponent $z=2$. 
The interaction $g$ has a canonical scaling dimension $[g]=3-d/2$. This theory has an upper critical dimension of $d=6$. In Eq.(\ref{Yukawa}), we have only kept the most relevant interaction term $\mathcal{L}_I$ for our discussions of the strong-coupling fixed-point physics when $\lambda<0$. 

Let us first focus on the case $\Gamma=\tau_x\sigma_y$. Later, we will show that such Yukawa interaction $g\psi^T\tau_x\sigma_y\psi\phi$ always leads to the condensation of the scalar field, i.e., $\langle\phi\rangle=\phi_0\neq0$ in 2D. This generates a term equivalent to an $s$-wave pairing with the same phase as $d_{xy}$. The fixed-point Hamiltonian then becomes
\begin{equation}\label{nodes}
\mathcal{H}_\text{fp}=\frac12\psi^T[\tau_x\sigma_y(d_{xy}+2g\phi_0)+\tau_z\sigma_y d_{x^2-y^2}]\psi.
\end{equation}
Without loss of generality, we take $g\phi_0<0$. The energy spectrum is gapless with two Dirac cones at $K_\pm=\pm(k_0, k_0)$ with $k_0=\sqrt{|g\phi_0|}$. In analogy to graphene, we call $K_\pm$ two valleys. 
Let us introduce Pauli matrices $\eta_{x,y,z}$ for the valley index and define $\Psi^T=(\psi^T_+,\psi^T_-)$ with the subscript $\pm$ labeling the two valleys. By expanding the Hamiltonian near each valley to linear order and rotating the coordinate by $-\pi/4$, we obtain the effective Hamiltonian with $N_f^B=1$ and $z=1$,
\begin{equation}\label{effective}
\mathcal{H}=\frac12\Psi^T[\eta_y\tau_x\sigma_y v(-i\partial_y)+\eta_y\tau_z\sigma_y v(-i\partial_x)]\Psi.
\end{equation}

This is a rare example where both the degrees of freedom  $N^B_f$ and dynamic critical exponent $z$ explicitly rely on interactions in addition to protecting symmetry $G_p$, i.e.,
\begin{eqnarray}
N^B_f=N^{B}_f(G_p; \lambda),\quad z=z(G_p;\lambda),
\label{degrees2}
\end{eqnarray}
which is a unique feature of tQCPs. To the best of our knowledge, such a particular scenario rarely appears in the discussions of conventional QCPs. In the view of Eq. (\ref{degrees2}), both the dynamic critical exponent $z$ and the degrees of freedom at tQCPs can be more appropriately considered as {\em emergent properties of tQCPs} rather than {\em fundamental input parameters} required to identify the universality of quantum critical phenomena. They can be viewed as the direct consequences of quantum dynamics of the emergent scalar field $\phi$ \cite{MC}. 

\begin{figure}
    \centering
    \includegraphics[width=\columnwidth]{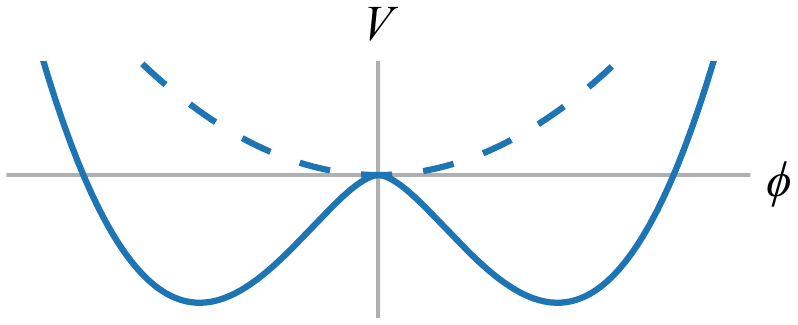}
    \caption{The solid line is the one-loop effective potential $V(\phi)$ of the emergent bosonic field for a bare mass $M^2 (> 0)$, and arbitrary coupling $g$ in (2+1)D. Notice that we have $\langle \phi \rangle \neq 0$, implying a spontaneous breaking of the parity symmetry 
    at any coupling strength $g$ and arbitrary mass $M^2$. The dashed line is the tree level potential where the parity symmetry remains unbroken. }
    \label{potential}
\end{figure}

To show the condensation of $\phi$, we compute the one-loop effective potential of $\phi$,
\begin{equation}
\begin{split}
V&=\frac12M^2\phi^2+\int\frac{d\omega}{2\pi}\int\frac{d^2k}{(2\pi)^2}\text{tr}\sum_{n=1}^\infty\frac1{n}[g\phi\tau_x\sigma_yG(\omega,{\bf k})]^{n},\\
&=\frac12M^2\phi^2-\frac1{16\pi}g^2\phi^2\left[\alpha+\ln\left(\frac{(2\Lambda_0^2)^2}{g^2\phi^2}\right)\right], 
\end{split}
\end{equation}
where $G(\omega,{\bf k})=1/(i\omega-H_0)$ is the Green's function for massless fermions, $H_0=\tau_x\sigma_y (2k_xk_y)+\tau_z\sigma_y (k_x^2-k_y^2)$, $\alpha\approx3.386$ and $\Lambda_0$ is the momentum cutoff.
By minimizing the effective potential, we find it has two minima at 
\begin{equation}
    \phi_0=\pm\phi_c\exp\left\{-\frac{4\pi M^2}{g^2}\right\},
\end{equation}
where $\phi_c=\frac{2\Lambda_0^2}{g}\exp{\{(\alpha-1)/2\}}$.
The one-loop effective potential is shown in Fig. \ref{potential} \cite{data}. This illustrates the strong-coupling physics indicated in Fig.\ref{phase}. 

When we take into account the normal state dispersion $\epsilon_k$, the fixed-point Hamiltonian Eq. (\ref{nodes}) becomes
\begin{equation}
\mathcal{H}_\text{fp}=\frac12\psi^T[\tau_x\sigma_y(d_{xy}+2g\phi_0)+\tau_z\sigma_y d_{x^2-y^2}+\tau_y(\epsilon_k-m)]\psi.
\end{equation}
The energy spectrum remains gapless with two Dirac cones. The only difference is that the tQCP is shifted to $m=\epsilon_k$, where $k=\sqrt2 k_0$. One can again expand the fixed-point Hamiltonian near each valley and the resultant effective Hamiltonian is still of the Dirac form with $N_f^B=1$ and $z=1$.

Similarly, if $\Gamma=\tau_z\sigma_y$, the Yukawa interaction $g\psi^T\tau_z\sigma_y\psi\phi$ also leads to the condensation of $\phi$ field, which in turn generates a term equivalent to an $s$-wave pairing with the same phase as $d_{x^2-y^2}$. Following the same analysis, we again have an effective Hamiltonian with $N_f^B=1$ and $z=1$.

In addition to the change of dynamic critical exponent and degrees of freedom, we would like to mention another possibility. The Yukawa interaction can generate a dynamical mass analogous to the famous Gross-Neveu model \cite{GrossNeveu}. This happens for $\Gamma=\tau_y$. For simplicity, we again neglect $\epsilon_k$, which does not affect the result qualitatively. At $m=0$, the renormalized effective potential becomes
\begin{equation}
V=\frac12 M^2\phi^2+\frac{g^2\phi^2}{8\pi}\left(\ln{\frac{\phi^2}{\phi_c^2}-3}\right),
\end{equation}
with minima at
\begin{equation}
\phi_0=\pm\phi_c \exp\left\{-\frac{2\pi M^2}{g^2}\right\},
\end{equation}
where $\phi_c=\frac{2\Lambda_0^2}{g}$.
The condensation of $\phi$ generates a dynamical mass term $m_{dyn}=2g\phi_0$ and the tQCP becomes gapped
\begin{equation}    \mathcal{H}_\text{fp}=\frac12\psi^T(\tau_x\sigma_yd_{xy}+\tau_z\sigma_yd_{x^2-y^2}+\tau_y m_{dyn})\psi.
\end{equation}
In this case, the superconducting gap never closes. On the other hand, the state topology is different for $m>0$ and $m<0$. Therefore, the original continuous phase transition at the tQCP becomes an interaction induced first order phase transition.

\section{Discussions and Outlook}
In this paper, we discussed the important effect of interactions in defining the universality of tQCPs. In contrast to conventional QCPs whose universality is defined by symmetry and dimensionality alone for a given dynamic critical exponent, the universality of tQCPs can further strongly depend on interactions. Namely, both the dynamic critical exponent $z$ and the degrees of freedom of tQCPs can be varied by tuning interactions. When defining the universality of tQCPs, one needs to specify not only the protecting symmetry and the change of topological invariants but also the interaction. 

Finally, we would like to comment on a possible connection between the tQCPs studied in this paper and fractional quantum Hall states. The $z=2$ gapless tQCPs have identical correlations as the Haldane-Rezayi (HR) states \cite{HaldaneRezayi,Read2000}. 
In many discussions of non-Abelian fractional quantum Hall effect, various Pfaffians constructed out of a two-particle correlation $g(Z)$ have been conveniently introduced to define states.
The two-particle correlation $g(Z)$ for the HR state with $Z=x+iy$ scales as
\begin{eqnarray}
g(Z) \sim \frac{1}{Z^2},
\end{eqnarray} 
which is also the characteristic of $z=2$ gapless tQCPs.
By contrast, the $z=1$ gapless tQCPs have a distinctly different property characterized by 
\begin{eqnarray}
g(Z) \sim \frac{1}{Z|Z|}\sin[\text{Re} (\bar{k}_0\cdot Z)];\quad \bar{k}_0 =k_{0x}-ik_{0y}.
\label{z1FQHE}
\end{eqnarray} 
Furthermore, Eq. (\ref{z1FQHE}) is also different from the Pfaffian factor in a gapless critical state of Moore-Read Paffian state, $g(Z) \sim \frac{1}{Z|Z|}$ \cite{MooreRead}. The correlation in Eq. (\ref{z1FQHE}) can be applied to construct an incompressible state that also breaks the translational and rotational symmetries. This can be considered as an example of {\em incompressible} stripes, an interesting future direction to further pursue.

In this paper, we have focused on local interactions so that the effective field theories are local. When long-range interactions are concerned, critical points may only have a scale symmetry rather than the full conformal symmetry, and the effective field theory can be nonlocal. It is worth noting that for conventional QCPs with long-range interaction, the dynamic critical exponent may be affected by the range (but {\it not} the strength or other details) of the interaction. An example of ferromagneitc XXZ spin model with power-law interactions $-J/r^\alpha$ was studied previously, where the dynamic critical exponent is affected by the decay power $\alpha$ \cite{Yang19}.

In this paper, we have only focused on internal or nonspatial symmetries, which is standard in the classification of topological insulators and superconductors. It is possible that in addition to topology, crystalline symmetries may also constrain the value of dynamic critical exponents in the free sector of the theory.

\appendix

\section{Strong-coupling fixed point}
In addition to the Lagrangian
\begin{eqnarray}
&\mathcal L_\psi=\frac12\psi^T(\partial_\tau+H_0)\psi,\nonumber\\
&\mathcal L_\phi=\frac12[(\partial_\tau\phi)^2+(\nabla^2\phi)^2+M^2\phi^2],\nonumber\\
&\mathcal L_{I}=g\psi^T\Gamma\psi\phi, \label{sm:yukawa}
\end{eqnarray}
there also exist scalar field self-interactions
\begin{eqnarray}
    \mathcal{L'}=u_4\phi^4+u_6\phi^6+...,
\end{eqnarray}
i.e., terms of the form $u_{2n}\phi^{2n}$ with $n\geq2$. The canonical scaling dimensions of interactions are $[g]=3-d/2$ and $[u_{2n}]=2(n+1)-(n-1)d$.  
In the following, we illustrate: (1) the free particle theory with $\mathcal{L}_I=\mathcal{L}'=0$ is an unstable fixed point, and the interacting theory always flows to a strong-coupling phase;
(2) the anomalous dimensions generated by the Yukuwa interaction $\mathcal{L}_I$ can turn $u_{2n}$ irrelevant. 

Let us first write down the one-loop RG equations for the model in the gapless limit in spatial dimension $d<6$,
\begin{equation}
    \gamma=\frac{1}{2}\frac{d\ln Z}{d\ln\Lambda}=\frac{\Omega_{d}}{(2\pi)^d}\frac{\tilde g^2}{ c_\Gamma},\\
\end{equation}
\begin{equation}
    \beta(g^2)=\frac{d\tilde g^2}{d\ln\Lambda}=-[(6-d)-2\gamma]\tilde g^2,\\
\end{equation}
\begin{equation}
\begin{split}
    \beta(u_{2n})=\frac{d\tilde u_{2n}}{d\ln\Lambda}=&-[2(n+1)-(n-1)d-2n\gamma]\tilde u_{2n}\\
    &+a_{2n}\tilde u_4\tilde u_{2n}+b_{2n}\tilde g^{2n},
\end{split}
\end{equation}
where $\Omega_d$ is the solid angle of the $(d-1)$-sphere, $Z$ is the field renormalization, $\tilde g^2=g^2/\Lambda^4$, $\tilde u_{2n}=u_{2n}/\Lambda^4$, $a_{2n}$, $b_{2n}$ are numerical coefficients with $a_{2n}$ being positive, and $c_\Gamma$ is a numerical coefficient depending on the choice of $\Gamma$.

Near the noninteracting fixed point $\tilde g^2=\tilde u_{2n}=0$, $n\geq2$, $\mathcal{L}_I$ and $\mathcal{L}'$ are both relevant operators, implying the instability of free fixed point and a strong-coupling phase.
By setting $\beta(g^2)=0$, we obtain an IR stable fixed point with $\gamma=3-d/2$ and $\tilde g_c^2=(3-d/2)(2\pi)^dc_\Gamma /\Omega_d$. By substituting $\gamma=3-d/2$ into $\beta(u_{2n})$, the scaling dimensions of $u_{2n}$ become negative $[u_{2n}]=2+d-4n<0$ for $d<6$. Thus, the Yukawa interaction is the most relevant interaction.

Indeed, we find that the solution to the above RG equations always flows to a strong-coupling fixed point. To explicitly illustrate the solution, without loss of generaity, we take the large-$N$ limit to simplify the RG equations.
We fix $g^2N=w$, and $\beta(u_{2n})$ becomes
\begin{equation}
    \beta(u_{2n})=-(d+2-4n)\tilde u_{2n}+a_{2n}\tilde u_4\tilde u_{2n},
\end{equation}
where we have substituted in $\gamma=3-d/2$ and the terms $b_{2n}\tilde g^{2n}$ in $\beta(u_{2n})$ are of order $O(\tilde w^nN^{1-n})$, thus vanishing in the large-$N$ limit.

Let us first look at $\beta(u_4)$,
\begin{equation}
    \beta(u_4)=(6-d)\tilde u_4+a_4\tilde u_4^2.
\end{equation}
It has a noninteracting IR stable fixed point $\tilde u_{4c}=0$, different from the standard Wilson-Fisher physics. Plugging this into $\beta(u_{2n})$, $n\geq3$, we have
\begin{equation}
    \beta(u_{2n})=-(d+2-4n)\tilde u_{2n},\end{equation}
whose IR stable fixed point is $\tilde u_{2n,c}=0$. Therefore, all $u_{2n}\phi^{2n}$, $n\geq2$ terms are irrelavent. In this limit, the strong-coupling fixed point is simply given as $\tilde g_c^2=(3-d/2)(2\pi)^dc_\Gamma /\Omega_d$ and $\tilde u_{2n,c}=0$ for $d<6$.

\section{Spontaneous symmetry breaking in 2D}
Next, we turn to the $d=2$ case, where we show that the condensation of $\phi$ always occurs as suggested by the RG equations.
The RG equation for $M^2$ is 
\begin{eqnarray}
    \beta(M^2)&=\frac{d\tilde M^2}{d\ln\Lambda}=-(4-2\gamma)\tilde M^2+\frac{4}{\pi c_\Gamma}\tilde g^2,
\end{eqnarray}
where $\tilde M^=M^2/\Lambda^4$.
For $d=2$, we have $\gamma=2$ and $\tilde g_c^2=4\pi c_\Gamma$ at the IR stable strong-coupling fixed point. Plugging these into $\beta(M^2)$, we have  
\begin{equation}
    \beta(M^2)=16>0,
\end{equation}
which suggests that $\tilde M^2$ always flows to $-\infty$ in the IR limit. The negative value of $M^2$ in the IR limit suggests the condensation of $\phi$, which is in agreement with our calculation of the effective potential in (2+1)D in the main text.

\begin{acknowledgments}
F.Z. wants to thank Duncan Haldane, John McGreevy, Ashvin Vishwanath, and Xiao-Gang Wen for inspiring discussions. We are also thankful to the support and hospitality of the winter workshop on {\em Quantum Criticality and Topological Phase Transition}  at the Croucher Advanced Study institute and the Chinese University of Hong Kong in December 2023 where part of the current research on tQCPs was initiated. This project is in part supported by an NSERC (Canada) discovery grant under grant number RGPIN-2020-07070. ZC Gu is supported by funding from Hong Kong's Research Grants Council (RGC Research Fellow Scheme 2023/24, No. RFS2324-4S02). F.Y. is supported by the Start-up Research
Fund of Southeast University (RF1028624190).
\end{acknowledgments}


\begin{thebibliography}{References}
\bibitem{MC} If we treat the intrinsic emergent dynamic scalar field $g\phi$ induced by interactions as an external static tuning parameter $F$,
then $F\neq 0$ maps out the strongly coupling $z=1$ gapless phase while $F=0$ maps out the whole weakly coupling $z=2$ gapless phase.
From the point of view of an external parameter $F$, the $N^{B}_f=\frac{1}{2}$ fundamental fermonic representation with $z=2$ dynamics appears at a tricritical point which terminates the $z=1$ gapless phase doubling the number of fundamental fermionic fields for $G_p=SU(2)$. However, the view of dynamic scalar fields $\phi$ is more general
as $F$ can be treated as the meanfield limit of the dynamics. It is also more appropriate and generic as for a very wide class of weakly interacting $C$ class states, we anticipate that $\langle \phi \rangle = 0$ which naturally leads to a well defined gapless phase with $z=2$.
\end{thebibliography}
\end{document}